\documentclass[10pt,
twocolumn,
showpacs,
superscriptaddress,
floatfix,
longbibliography,
aps,
pra,
reprint]
{revtex4-2}

\usepackage[english]{babel}

\usepackage{physics}

\usepackage{amsmath}

\usepackage{mathtools}

\usepackage{amssymb}

\usepackage{graphicx}

\usepackage[colorlinks=true, allcolors=blue]{hyperref}

\begin{document}

\title{The effects of shot noise on the quantum computation of NMR spectra}
\author{Sebastian Walch}
\affiliation{Institut f\"ur Theoretische Physik and Center for Integrated Quantum Science and Technology, Universit\"at T\"ubingen, Auf der Morgenstelle 14, 72076 T\"ubingen, Germany.}
\author{Keith R. Fratus}
\author{Jan-Michael Reiner}
\affiliation{HQS Quantum Simulations GmbH, Rintheimer Stra\ss e 23, 76131 Karlsruhe, Germany.}
\author{Igor Lesanovsky}
\affiliation{Institut f\"ur Theoretische Physik and Center for Integrated Quantum Science and Technology, Universit\"at T\"ubingen, Auf der Morgenstelle 14, 72076 T\"ubingen, Germany.}
\affiliation{School of Physics and Astronomy and Centre for Mathematics and Theoretical Physics of Quantum Non-Equilibrium Systems, The University of Nottingham, Nottingham, NG7 2RD, United Kingdom.}

\begin{abstract}
       Recent advances in the field of quantum computing hardware motivate the search for applications which demonstrate so-called quantum advantage. One promising use case that has been identified is the simulation of quantum many-body systems. The computational resources required for performing such a simulation on a classical computer generally grow exponentially with the size of the system being modeled, which is ultimately due to the exponential growth of the Hilbert space in which the dynamics of such a system take place. While digital quantum computers natively evolve quantum states directly in such a Hilbert space, thus naively avoiding this problem, the result of such a computation is typically not obtained as a deterministic output. Rather, it requires projective measurements which are fundamentally affected by shot noise. Any desired expectation values must therefore be reconstructed from repeated measurements, making the number of those measurements a relevant computational resource, and thus an important consideration for any potential claims of quantum advantage. In this work, we study how this resource scales with system size (a scaling which itself depends on the desired accuracy of the final result), using the simulation of \textit{nuclear magnetic resonance} (NMR) spectra as a test case. We study this scaling for both real-world molecules, as well as a class of model NMR Hamiltonians which allow for efficient large-scale simulations with one-dimensional, two-dimensional, and all-to-all interactions. We find that the required resources increase only weakly with molecular size, far below the exponential growth of the underlying Hilbert space. This result suggests that shot noise should not pose a fundamental barrier to achieving quantum advantage in the simulation of NMR systems, and perhaps for many-body systems more broadly.

\end{abstract}

\maketitle


\clearpage
\section{Introduction} One central promise of quantum technology is the notion of quantum advantage, which is the idea that a quantum device could provide a computational benefit over available classical methods for a given task. Such an advantage could manifest in several ways, for example in terms of runtime, memory requirements, or by offering better scalability than an algorithm running on a classical computer~\cite{Shor_94, Loyd_96, Grover_96, Shor_97, Nielsen_10, Zoller_11, Wendin_17, Zhang_17, Preskill_18, Krantz_19, Bruzewicz_19, Bauer_20, Sels_20, McArdle_20, Cerezo_21, Kiani_22, Noiri_22, Seetharam_23, Fauseweh_24, Blekos_24, Marthaler_25, Fratus_25, Google_25}. 

\begin{figure*}[t!]
     \centering
     \includegraphics[width = 17.92cm]{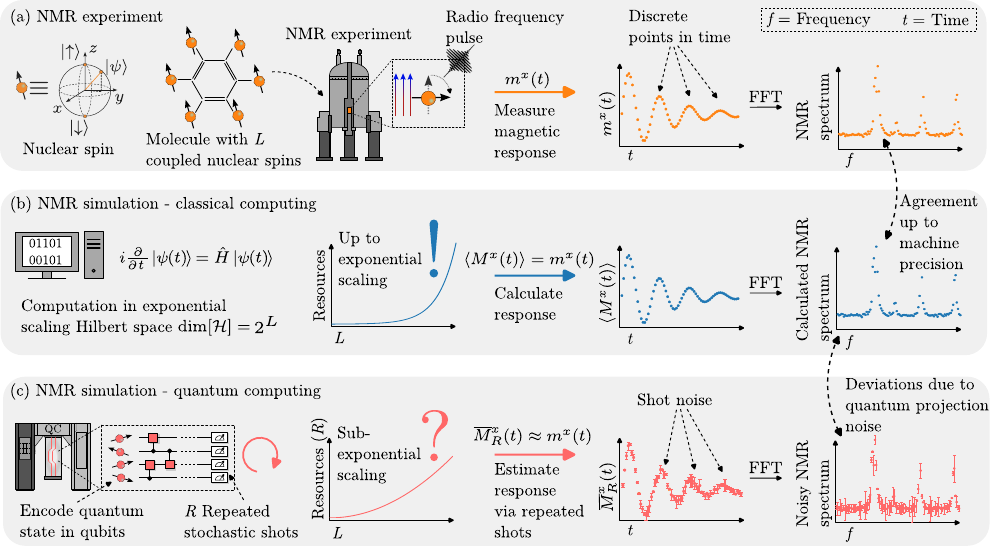}
     \caption{\textbf{NMR spectroscopy and computational resources.} (a) Molecules containing $L$ NMR-active nuclei represent a many-body system modeled by coupled quantum spins [cf.~Eq.~\eqref{eq:heisenberg_hamiltonian}]. In an external magnetic field, radio-frequency pulses induce spin dynamics whose time-dependent magnetic response $m^x(t)$ (and $m^y(t)$) is measured. The corresponding NMR spectrum is obtained via a \textit{fast Fourier transform} (FFT) of this signal. (b) NMR spectra may be computed up to machine precision with classical computers, for example from the FFT of the expectation values of magnetization operators $\langle \hat{M}^x(t)\rangle$ (and $\langle \hat{M}^y(t)\rangle$). However, since the Hilbert space grows exponentially with the number of NMR-active nuclei $L$, the required memory and runtime generally also increase exponentially. (c) The nuclear spins map naturally onto the qubits of a quantum computer, allowing the many-body dynamics to be simulated through the application of quantum gates on a qubit register. The time evolution of the magnetization $\langle \hat M^x(t)\rangle$ is approximated by the noisy quantity $\overline{M}^x_R(t)$ obtained from repeated measurements (shots). This fundamental property of digital quantum computers introduces shot noise into the simulated time signal and, consequently, into the spectrum. The goal of our study is to quantify the number of required shots to achieve a desired overlap between noisy spectra reconstructed from repeated shots with spectra obtained from numerically exact simulations as a function of $L$. In the following, this overlap will be denoted as the accuracy of the noisy spectra.}
    \label{fig:introduction}
 \end{figure*}

One broad area which has been proposed as a key candidate for achieving quantum advantage is the simulation of many-body quantum systems, especially quantum spin systems~\cite{Loyd_96, Sels_20, Seetharam_23, Fauseweh_24, Marthaler_25, Fratus_25}. These systems are relevant in many areas of physics, chemistry, and materials science. Their exact treatment on classical computers is generally challenging due to the exponential growth of the Hilbert space [cf.~Fig.~\ref{fig:introduction}(b)]. In particular, storing and manipulating the full many-body quantum state requires computational resources that grow exponentially with system size. A digital quantum computer offers a different route, since the many-body quantum state may be encoded in a register of qubits and manipulated through unitary gates.

However, the encoded quantum state cannot be read out directly - information must be extracted via projective measurements of the qubit register. Relevant quantities such as expectation values must therefore be estimated from repeated measurements, so-called shots. Since each shot yields a stochastic outcome, a finite number of shots leads to statistical fluctuations of the estimated quantities, known as shot noise~\cite{Nielsen_10}. This is inherent to the operation of quantum computers, but can be mitigated by increasing the number of shots, which therefore becomes an important resource for quantum computing applications. It is essential to understand how the number of shots that is required to achieve a desired accuracy scales with the size of the simulated many-body system.

We study this question in the context of \textit{nuclear magnetic resonance} (NMR) spectra~\cite{Levitt_08, Keeler_10, Kuprov_11, Clayden_12, Hosur_22, Fratus_25}. NMR is a highly relevant technique in chemistry and medical research, where it is widely used, for example, in drug discovery~\cite{Claridge_16, Breitmaier_02, Pellecchia_08, Harner_13}. In an experiment, NMR spectra are obtained from the magnetic response of molecules to radio-frequency pulses [cf.~Fig.~\ref{fig:introduction}(a)]. NMR molecules may be modeled as coupled nuclear spin systems with Heisenberg interactions~\cite{Levitt_08, Keeler_10}, whose spin-1/2 degrees of freedom map in a natural way to the qubits of a digital quantum computer. The simulation of NMR spectra on a quantum computer entails computing the time evolution of the quantum states of molecules and calculating expectation values of magnetization operators. Fourier transforming these time dependent magnetization expectation values yields the NMR spectrum. This has been the subject of several works on digital quantum computing~\cite{Sels_20, Seetharam_23, Marthaler_25, Burov_25}. Independently of whether the computation of NMR spectra with quantum computers ultimately leads to a quantum advantage~\cite{Fratus_25}, NMR Hamiltonians nevertheless provide a concrete and relevant setting in which the effects of shot noise from repeated measurements may be studied. 

In this work, we investigate how the accuracy of NMR spectra -- computed from the time-evolution of spin dynamics on quantum computers -- depends on the number of shots. The corresponding setting is summarized in Fig.~\ref{fig:introduction}(c). All other sources of error, such as decoherence due to imperfect gates or coupling to an environment, as well as approximations of the time evolution via Trotterization~\cite{Shor_95, Steane_95, Nielsen_10, Fowler_12, Lidar_13, Terhal_15, Krinner_22}, are deliberately neglected, such that statistical fluctuations due to the finite number of shots constitute the only source of uncertainty. To enable this analysis for large molecules with up to 50 NMR-active nuclei, we employ a simplified NMR model allowing access to numerically exact reference dynamics [cf.~Eq.~\eqref{eq:ising_hamiltonian}]. Throughout, we consider a low-field NMR setting and investigate different molecular geometries. 

As a central result, we find that the number of shots does not scale exponentially with system size. Simulations of the full NMR dynamics [cf.~Eq.~\eqref{eq:heisenberg_hamiltonian}] for smaller systems suggest that the observed scaling is not specific to the simplified model and may apply generally.


\section{Nuclear magnetic resonance spectroscopy}

The central object governing the static and dynamical properties of molecules studied through NMR is the Heisenberg spin Hamiltonian
\begin{equation}
    \hat{H}_\theta = \sum_{l=1}^L h_l \hat{I}^z_l + \sum_{p<l}^L J_{pl}\mathbf{\hat{I}}_p \cdot \mathbf{\hat{I}}_l.
    \label{eq:heisenberg_hamiltonian}
\end{equation}
The operators $\hat{I}^\alpha$ with $\alpha\in\{x,y,z\}$ denote the components of the dimensionless spin-$1/2$ operator, i.e.\ $\hat{I}^\alpha=\hat{\sigma}^\alpha/2$, where $\hat{\sigma}^\alpha$ are the Pauli matrices. The indices $l$ and $p$ denote the $l$-th and $p$-th nuclei respectively and $L$ is the total number of NMR-active nuclei in the molecule. The first term of the Hamiltonian describes the Zeeman interaction of the nuclei with an externally applied magnetic field, quantified by the local Zeeman coefficients $h_l$. The second term represents the Heisenberg interaction between nuclei which is quantified by interaction strength $J_{pl}$.
An NMR molecule is thus characterized by the parameter set $\theta = \{h_l,J_{pl} \}$.

The way we compute NMR spectra is closely aligned with the actual experimental measurement procedure. In the following we therefore describe the experimental steps together with their computational counterparts. A sample of molecules is placed in a strong, constant magnetic field. This results in a net magnetization in the sample, i.e.\ nuclear spins tend to align with the external field. Here we take this external field to be in the $z$-direction. After letting the sample equilibrate, radio-frequency pulses rotate the spins of a given isotope (most often protons) into the $xy$-plane, where they begin to precess (in this work, we assume that all spins in the molecule are protons, and thus ignore the case of heteronuclear molecules). This precession, also known as the "free induction decay" (FID), generates a magnetic response in the detector coils of the experiment, which is then recorded. In the quantum-mechanical description, the average magnetic moment of a single molecule along a specified direction is given by the expectation value of the corresponding spin operator $\hat{M}^\alpha \equiv \sum_{l=1}^L  \hat{I}^\alpha_l$, which is given as
\begin{equation}
    \langle \hat{M}_\theta^\alpha(t)\rangle = \frac{1}{2^L}\text{Tr} \left [  \hat{M}_\theta^\alpha(t)  \hat{M}^x  \right ],
    \label{eq:correlation_function_gen}
\end{equation}
where $\hat{M}^x$ enters the equation through a high temperature expansion of the thermal density matrix~\cite{Fratus_25}.
Here, $\hat{M}_\theta^\alpha(t) = \hat{U}_\theta^\dag(t)\hat{M}^\alpha \hat{U}_\theta(t)$ denotes the operator $\hat{M}^\alpha$ time evolved via $\hat{U}_\theta(t)=e^{-2 \pi i t\hat{H}_\theta}$. The magnetic signal detected in the coils of an NMR experiment is proportional to the sum of a macroscopically large number $N_{\text{mol}}$ of such magnetic moments. Since we neglect the effects of inter-molecular interactions during the FID, this sum equals $N_{\text{mol}}$ times the expectation value of a single molecule, up to relative fluctuations on the order $1/\sqrt{N_{\text{mol}}}$. Therefore, the recorded magnetic response is, to a good approximation, proportional to the magnetization expectation value of a single molecule, Eq.~\eqref{eq:correlation_function_gen}~\cite{Levitt_08}. Furthermore, such a macroscopic measurement, with sufficiently low resolution compared to the scale of individual magnetic moments, can be effectively regarded as non-invasive~\cite{Poulin_2005}, allowing us to study the expectation value of a single molecule, without any consideration of any possible collapse dynamics.

In addition to this unitary evolution,  Eq.~\eqref{eq:correlation_function_gen}, the measured signal decays in time due to decoherence processes~\cite{Clayden_12, Hosur_22, Levitt_08, Keeler_10}. The simplest way to model this effect phenomenologically is by introducing exponential decay, which is characterised by a single so-called broadening parameter $\eta$~\cite{Levitt_08, Keeler_10, Clayden_12, Sels_20, Hosur_22, Fratus_25},
\begin{equation}
    \langle \hat{M}_\theta^\alpha(t)\rangle
    \rightarrow
    \langle \hat{M}_\theta^\alpha(t)\rangle e^{- \eta \pi t}.
\end{equation}
This approximation assumes a homogeneous decay rate for all spins, which is valid when the decay of the signal is dominated by a single relaxation timescale and accurate for NMR calculations.

The results of an NMR experiment are usually interpreted through the complex combination
\begin{equation}
S_\theta(t_n) =
\left(
\langle \hat{M}_\theta^x(t_n)\rangle
+ i\langle \hat{M}_\theta^y(t_n)\rangle
\right)
e^{-\eta \pi t_n}.
\label{eq:FID}
\end{equation}
For the purposes of this work, when we refer to the FID, we will usually mean this specific complex combination (even though this usage is not strictly consistent with the conventional notation). The signal in the coils is necessarily recorded with a finite sampling rate. We assume equidistant sampling times $t_n=n\Delta t$ with $n=0,1,\dots,N-1$, giving $N$ time points separated by $\Delta t$ [see Fig.~\ref{fig:introduction}(a)]. The corresponding NMR spectrum is then computed from the discrete time signal using a \textit{fast Fourier transform} (FFT)
\begin{equation}
    S_\theta(f_k)
    =
    \operatorname{Re}
    \left[
        \sum_{n=0}^{N-1}
        S_\theta(t_n)
        e^{-2\pi i f_k t_n}
    \right],
    \label{eq:spectrum}
\end{equation}
where the spectrum is evaluated on the $N$ discrete frequencies
\begin{equation}
  f_k = \frac{k}{N\Delta t}, \qquad
  k \in \left\{-\tfrac{N}{2},\dots,\tfrac{N}{2}-1\right\}.
  \label{eq:freqs}
\end{equation}
Accordingly, throughout the following considerations, time-domain signals are treated as discrete sequences and their spectra as discrete frequency-domain representations.

The shape of the spectrum is determined by the local Zeeman coefficients and is further modified by the spin-spin interactions. In the absence of interactions, where all coupling strengths $J_{pl}$ are set to zero, each NMR-active nucleus generates one Lorentzian peak of linewidth $\eta$. This peak is centered at the corresponding precession frequency of that nucleus, determined according to the local Zeeman coefficients $h_l$. The interactions between the nuclei typically split these peaks into multiple components~\cite{Levitt_08, Keeler_10, Clayden_12, Hosur_22}, thereby increasing the complexity of the spectra. When the spectra consist of comparatively well-separated resonances and simple multiplet structures, these features can frequently be assigned directly to local Zeeman coefficients and a small number of spin-spin couplings.

When multiple nuclei contribute to resonances at similar frequencies and the interaction-induced splittings from several couplings -- including distance-dependent couplings -- overlap significantly, the direct interpretation then becomes more difficult. In this regime, simulations become valuable for interpreting experimental spectra. Instead of assigning spectral features directly, one chooses a set of Hamiltonian parameters $\theta$ and computes the corresponding spectrum $S_\theta(f_k)$, which can then be compared to the measured one~\cite{Kuprov_11, Fratus_25}. For generic NMR molecules, no closed-form expression for $S_\theta(f_k)$ exists. Consequently, calculating the spectrum requires solving the underlying many-body spin dynamics, for example by simulating the time evolution of the system.

The central computational task is therefore the evaluation of Eq.~\eqref{eq:correlation_function_gen}, i.e.\ the time evolution of the many-body spin system and the subsequent computation of magnetization expectation values. On a classical computer this can, in principle, be done up to machine precision by explicitly representing the quantum state or the relevant operators in the Hilbert space $\mathcal{H}$. However, for $L$ spin-$1/2$ particles this space has dimension $\dim[\mathcal{H}]=2^L$, so the required memory and runtime generally grow exponentially with system size. This makes exact simulations of NMR spectra prohibitive for molecules with more than $L\sim 25$ spins, as illustrated in Fig.~\ref{fig:introduction}(b). Such system sizes are common in NMR-relevant applications, making this exponential scaling a significant computational bottleneck. Approximate classical methods can mitigate this problem by trading accuracy for a more favorable scaling with $L$~\cite{Kuprov_11}.
In fact, for the most common NMR protocols performed on molecules in liquid solution, this accuracy tradeoff is negligible in practice, even for extreme parameter settings~\cite{Fratus_25}. Nevertheless, 1D liquid NMR still provides a concrete, relevant setting for understanding the effects of shot noise on a quantum computation. Ruling out the possibility of problematic shot noise scaling (which would pose a serious impediment to quantum advantage) in such directly accessible cases, before extending such an analysis to more computationally demanding many-body problems, is of course prudent.


\section{NMR spectra with shot noise}
\label{sec:shot_noise}

A quantum computer natively simulates many-body quantum dynamics. As described above, the relevant degrees of freedom of the NMR molecules are nuclear spins, which map naturally to the qubits of a quantum computer. The many-body quantum state can therefore be directly represented in the state of the qubit register. The unitary time evolution of NMR molecules is implemented as a gate sequence~\cite{Loyd_96, Nielsen_10, Sels_20, Seetharam_23, Fauseweh_24}. For the readout of the qubit register, projective measurements are performed. Each measurement -- a so-called shot -- yields exactly one eigenvalue of the corresponding measurement operator. Expectation values are estimated by averaging over the outcomes of these shots. This is illustrated in Fig.~\ref{fig:introduction}(c)~\cite{Nielsen_10, Sels_20, Seetharam_23, Burov_25}.

Later, we present how the expectation values in Eq.~\eqref{eq:correlation_function_gen} are estimated via repeated shots with a quantum computer. One shot entails three main steps: (i) preparation of an initial state, (ii) unitary time evolution, and (iii) a final projective measurement which yields an eigenvalue of the measured observable.  In the following, we detail each of those steps:

(i) Eq.~\eqref{eq:correlation_function_gen} may be interpreted as the expectation value of the operator $\hat{M}_\theta^\alpha(t)\hat{M}^x$ in the maximally mixed state $\rho_0=\frac{\mathbb{I}}{2^L}$. We now sample the initial states from $\rho_0$ by choosing eigenstates of $\hat M^x$. We represent these eigenstates as product states in the local $x$-basis. These eigenstates satisfy
\begin{equation}
\hat M^x\ket{x_j}=\lambda_j^x\ket{x_j}.
\label{eq:rho_x_basis}
\end{equation}
The eigenvalues $\lambda_j^x$ lie in the range $[-L/2,L/2]$ and are generally highly degenerate, since different product states $\ket{x_j}$ can yield the same total magnetization. The index $j$ represents one of the $2^L$ product states. This allows us to express the initial state as
\begin{equation}
\rho_0 =\frac{1}{2^L}
\sum_{j=1}^{2^L}
\ket{x_j}\bra{x_j}.
\end{equation}

(ii) The sampled initial states are then evolved in time using the unitary time-evolution operator generated by the Hamiltonian in Eq.~\eqref{eq:heisenberg_hamiltonian},
\begin{equation}
\hat{U}_\theta(t)\ket{x_j}
=
e^{-2\pi i t\hat{H}_\theta}\ket{x_j}.
\end{equation}
On a digital quantum computer this is typically implemented as a sequence of local gates. This often entails approximation errors, such as Trotter errors. We assume here that the time evolution is perfectly implemented.

(iii) The time-evolved state $\hat U_\theta(t)\ket{x_j}$ is measured projectively in the eigenbasis of $\hat M^\alpha$. This measurement yields an eigenvalue $\lambda_i^\alpha$, associated with the eigenstate $\ket{\alpha_i}$. For a given initial state $\ket{x_j}$, the corresponding probability is conditioned on $\ket{x_j}$ and reads
\begin{equation}
P^\alpha_\theta(i,t|j)
=
|\matrixel{\alpha_i}{\hat U_\theta(t)}{x_j}|^2 .
\end{equation}

In the following we discuss how repeated shots allow to estimate Eq.~\eqref{eq:correlation_function_gen}. To this end we use that it may be written as:
\begin{equation}
\langle \hat{M}_\theta^\alpha(t)\rangle
 = \frac{1}{2^L}
\sum_{j=1}^{2^L}
\sum_{i=1}^{2^L}
\lambda_j^x \lambda_i^\alpha
P^\alpha_\theta(i,t|j).
\label{eq:simple_correlation_function_x}
\end{equation}

We now sample the products $\lambda_j^x\lambda_i^\alpha$ appearing in Eq.~\eqref{eq:simple_correlation_function_x}. The value $\lambda_j^x$ is determined by the sampled initial state $\ket{x_j}$ from Eq.~\eqref{eq:rho_x_basis}, while the value $\lambda_i^\alpha$ is obtained from the projective measurement (which is in accordance with the conditional probability $P^\alpha_\theta(i,t|j)$). Thus, one shot realizes one value $(\lambda_j^x \lambda_i^\alpha)_r$, which depends on the Hamiltonian parameters $\theta$ through the probability distribution $P^\alpha_\theta(i,t|j)$. For a finite number of shots $R$, this gives the estimator
\begin{equation}
\overline{M}^\alpha_{R,\theta}(t)
=
\frac{1}{R}
\sum_{r=1}^{R} (\lambda_j^x \lambda_i^\alpha)_r
\xrightarrow{R\to\infty}
\langle \hat{M}_\theta^\alpha(t)\rangle .
\label{eq:finite_shot_estimator}
\end{equation}

In order to estimate the full signal Eq.~\eqref{eq:FID}, we sample $\overline{M}_{R,\theta}^x(t_n)$ and $\overline{M}_{R,\theta}^y(t_n)$ with $R$ shots each. The corresponding discrete NMR spectrum is then obtained from the estimated time signal via the FFT as 
\begin{equation}
        S_{R,\theta}(f_k) = \operatorname{Re}
    \left[
    \sum_{n=0}^{N-1}  \left ( \overline{M}_{R,\theta}^x(t_n) + i \overline{M}_{R,\theta}^y(t_n) \right ) w(n,k)\right]
    \label{eq:shot_noise_spectra}
\end{equation}
with $w(n,k)=e^{-\eta \pi t_n}e^{-2\pi i f_k t_n}$. As before, $t_n$ denotes the discrete time points at which the signal is estimated, which leads to the corresponding discrete frequencies $f_k$.

For finite $R$, the estimates $\overline            {M}^\alpha_{R,\theta}(t_n)$ fluctuate around the exact average value due to the stochastic nature of the repeated shots. These statistical errors enter the time-dependent signal at all discrete times $t_n$ and are propagated to the spectrum $S_{R,\theta}(f_k)$. The resulting spectrum is therefore a noisy approximation of the exact NMR spectrum, with deviations that decrease as the number of shots $R$ is increased.


\section{Scaling of required number of shots with system size}
\subsection{Setting and figure of merit}

Our goal is to quantify the scaling of the number of shots $R$ that is required to obtain NMR spectra Eq.~\eqref{eq:shot_noise_spectra} with a desired accuracy. Furthermore, we want to understand how the required number of shots scales with the number of NMR-active nuclei $L$.

In addition to shot noise, several other error sources may affect computations on a digital quantum computer. 
These consist of hardware errors, such as gate errors, where the applied gates do not exactly correspond to the desired gates. They also include decoherence effects due to the unwanted interaction of the qubits with their environment or other qubits, leading to a loss of information stored in the qubits over time. Further errors can arise from approximating the time evolution of the qubits, for example through Trotterization. These errors may in principle be mitigated or even corrected by employing quantum error correction. Shot noise, however, is more fundamental and will remain an obstacle in the era of error corrected quantum devices. We therefore focus on shot noise and assume that the resulting shot numbers provide a lower bound on the resources required to calculate high-accuracy NMR spectra.

The impact of shot noise on the estimation of NMR observables has been investigated in previous work. For example, Ref.~\cite{Sels_20} provides a worst-case bound for the variance of a single estimated magnetization value, scaling as $O(L^2/R)$. Here we are interested in computing NMR spectra from an NMR time series where the signal at each point in time is subject to shot noise. These errors are propagated by the FFT to the final obtained spectrum. We therefore quantify shot noise at the level of the reconstructed spectrum rather than at a single estimated time point.

To quantify the accuracy of a noisy spectrum $S_{R,\theta}(f_k)$, we compare its shape to the corresponding exact reference spectrum $S_{\theta}(f_k)$, defined in Eq.~\eqref{eq:spectrum}. To measure the overlap between these spectra, we use the cosine similarity

\begin{equation}
  C(S_{R,\theta}(f_k), S_\theta(f_k)) =
  \frac{\sum_{k=-N/2}^{N/2-1} S_{R,\theta}(f_k) S_\theta(f_k)}
       {||S_{R,\theta}(f_k)||\,||S_\theta(f_k)||}
        \label{eq:cosine_sim}
\end{equation}
with
\begin{align*}
  ||S_{R,\theta}(f_k)|| &\equiv \sqrt{\sum_{k=-N/2}^{N/2-1} (S_{R,\theta}(f_k))^2} \\
  ||S_\theta(f_k)||    &\equiv \sqrt{\sum_{k=-N/2}^{N/2-1} (S_\theta(f_k))^2}.
\end{align*}
All sums run over the full set of discrete frequencies of Eq.~\eqref{eq:freqs}. One advantage of the cosine similarity is that it does not depend on the overall normalization of either spectrum, a feature of the spectrum which carries no physical relevance to us. It is also invariant under a linear change of the frequency variable, such that it does not depend on the specific choice of units. This allows us to compare only the features that are directly relevant to the physical interpretation of the spectrum, for example the positions and widths of peaks. For the strictly positive spectra considered here, the cosine similarity lies in the range $[0,1]$, with $C(S_{R,\theta}(f_k),S_{\theta}(f_k))=1$ for identical spectral shapes.

We are interested in general statements about shot requirements, rather than the behavior of one specific molecule. We therefore consider an ensemble of molecules which is constructed by sampling Hamiltonian parameters $\theta_\nu$, where $\nu$ labels one explicit molecule with fixed local Zeeman coefficients and interaction strengths. For each sampled molecule $\theta_\nu$, we generate several independent noisy time series. We refer to each of these noisy time series as one run and label it by $\mu$. From each time series we then compute the noisy spectrum $S_{R,\theta_\nu}^{\mu}(f_k)$. Different runs for the same $\theta_\nu$ and $R$ correspond to different shot noise realizations and therefore yield different noisy spectra.
\begin{figure}[t!]
    \centering
    \includegraphics[width = 7.5 cm]{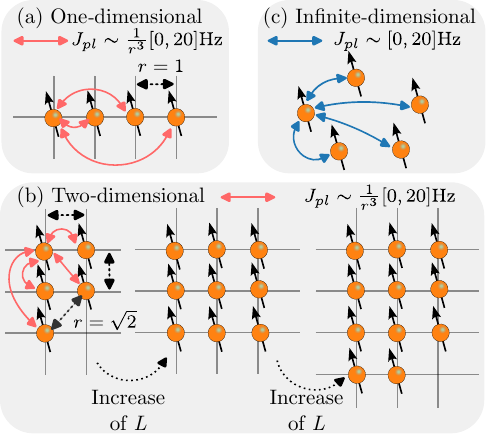}
    \caption{\textbf{Considered geometries.}{
    (a) One-dimensional chain with distance-dependent couplings $J_{pl}=K_{pl}/r_{pl}^3$.
    (b) Two-dimensional square lattice with the same distance dependence. For increasing $L$, the lattice is enlarged by completing square configurations and then adding spins around the boundary.
    (c) Infinite-dimensional geometry without spatial dependence, where $J_{pl}=K_{pl}$.
    For all geometries, $K_{pl}$ is sampled uniformly from $[0,20]\,\mathrm{Hz}$.}
\label{fig:geom}}
\end{figure}
The cosine similarity characterizes one noisy spectrum for one sampled molecule. It is convenient to work instead with its deviation from perfect agreement, which we refer to as the single-run infidelity,
\begin{equation}
  \mathcal{I}^\mu_\nu(R) =
  1 - C(S^\mu_{R,\theta_\nu}(f_k), S_{\theta_\nu}(f_k)),
\end{equation}
which vanishes when the noisy spectrum has exactly the shape of the reference spectrum and grows as the two shapes differ. To obtain a quantity that describes the typical spectral accuracy, we average it over the sampled system realizations $\theta_\nu$ and over the independent runs $\mu$ for each realization. This defines the infidelity
\begin{equation}
  \mathcal{I}(R,L) = \frac{1}{N_\mu N_\nu}
  \sum_{\nu=1}^{N_\nu} \sum_{\mu=1}^{N_\mu} \mathcal{I}^\mu_\nu(R),
  \label{eq:infidelity}
\end{equation}
where $N_\nu$ is the number of sampled system parameter sets and $N_\mu$ is the number of independent runs for each system.

For a sufficiently large number of shots $R$ the infidelity acquires the form
\begin{equation}
\mathcal{I}(R,L)
\approx
\frac{\mathcal{A}(L)}{R},
\label{eq:infidelity_scaling}
\end{equation}
which separates the dependence on the number of shots from the dependence on system size (see Appendix \ref{app:scaling} for the derivation). Increasing $R$ reduces the infidelity as $1/R$, while the remaining dependence on the system size $L$ is contained in the scaling function $\mathcal{A}(L)$ which determines how the number of shots required to reach a certain infidelity $\mathcal{I}$ increases with $L$. For example, if $\mathcal{A}(L)$ grew exponentially, the required number of shots would also grow exponentially with system size.

\subsection{Ising interactions}
\label{sec:ising_interaction}

We now determine the scaling function $\mathcal{A}(L)$ for different system sizes and investigate how it depends on the interaction geometry. This scaling analysis becomes computationally highly demanding as the system size increases. To some extent this difficulty may be overcome by employing approximate methods, for example tensor-network approaches or clustering methods~\cite{Fratus_25}. As a first step, however, we base the analysis on reference spectra that can be computed numerically exactly. To this end, we replace the Heisenberg interactions in Eq.~\eqref{eq:heisenberg_hamiltonian} by Ising interactions $J_{pl}\hat{I}^z_p\hat{I}^z_l$. This yields the Ising Hamiltonian
\begin{equation}
\hat{H}_\theta
=
\sum_{l=1}^{L} h_l \hat{I}_l^z
+
\sum_{p<l}^{L} J_{pl}\hat{I}_p^z\hat{I}_l^z ,
\label{eq:ising_hamiltonian}
\end{equation}
which is diagonal in the computational basis. We exploit this to compute the exact dynamics, and thereby the corresponding discrete FID and reference spectrum $S_{\theta_\nu}(f_k)$, for systems with up to $L=50$ spins. Later, we compare these results to smaller-scale simulations of the Heisenberg spin Hamiltonian Eq.~\eqref{eq:heisenberg_hamiltonian}.
For the Ising Hamiltonian, Eq.~\eqref{eq:ising_hamiltonian}, both the exact magnetization expectation value, Eq.~\eqref{eq:correlation_function_gen}, and higher moments of the magnetization can be computed numerically exactly. We exploit this to directly generate the noisy magnetization values entering the noisy spectra: For each Hamiltonian realization $\theta$, each discrete time $t_n$, and each direction $\alpha=x,y$, we compute the exact magnetization expectation value and the corresponding single-shot variance (see Appendix \ref{app:estimator} for the explicit expressions). We then draw $\overline{M}^\alpha_{R,\theta}(t_n)$ from the normal distribution that represents the average over $R$ shots. This reproduces the statistical effect of repeated measurements, while avoiding the explicit sampling of all individual shot outcomes, which would be infeasible for the large systems considered here. The noisy spectra are then computed according to Eq.~\eqref{eq:shot_noise_spectra}.

We consider a low-field NMR setting. In this regime, the local Zeeman coefficients (in a rotating frame) are expected to lie in the range of tens to hundreds of Hz and we sample them uniformly from $h_l\sim[20,200]\,\mathrm{Hz}$~\cite{Levitt_08, Keeler_10, Clayden_12, Hosur_22}. Note that such a low field strength is somewhat atypical in NMR experiments.

\begin{figure*}[t]
    \centering
    \includegraphics[width = 17.92cm]{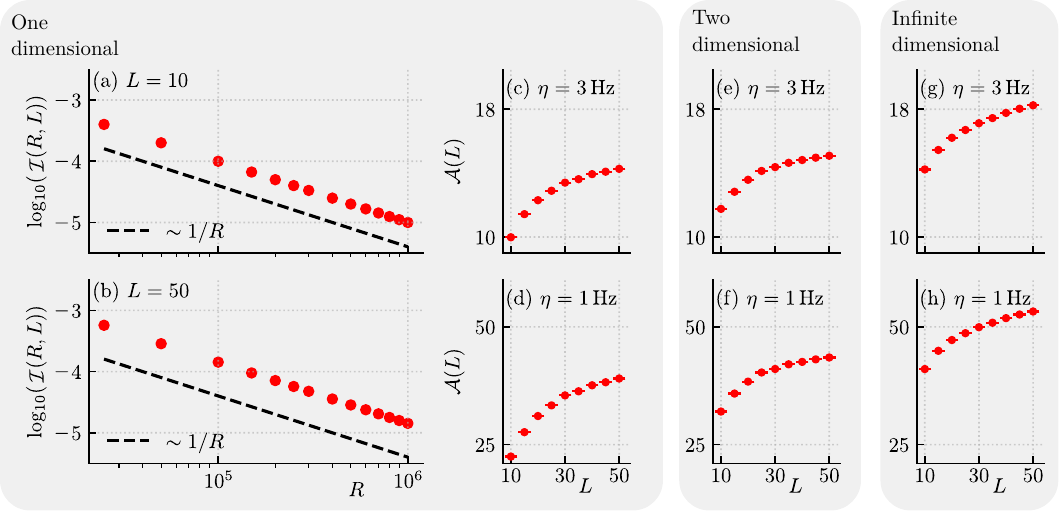}
    \caption{\textbf{Infidelity and scaling function.} (a,b) infidelity $\mathcal{I}(R,L)$ as a function of the number of shots $R$ for the one-dimensional geometry with broadening $\eta=3\,\mathrm{Hz}$, shown for (a) $L=10$ and (b) $L=50$. The scaling $\mathcal{I}(R,L)\sim R^{-1}$ is evident. (c)--(h) Scaling function $\mathcal{A}(L)$ for all considered geometries and broadening parameters. In all cases, $\mathcal{A}(L)$ increases with system size, indicating that larger molecules require more shots to achieve the same target infidelity. The growth becomes weaker for larger $L$. This flattening effect is most pronounced in the one- and two-dimensional geometries, while the infinite-dimensional geometry exhibits the strongest overall increase. Across all geometries, the broadening parameter $\eta$ does not affect this qualitative behavior of $\mathcal{A}(L)$.}
    \label{fig:result}
\end{figure*}
In contrast to the local magnetic-field terms, the spin-spin interactions depend on the spatial arrangement of the spins. We therefore consider three different interaction geometries, which are illustrated in Fig.~\ref{fig:geom}. In the one-dimensional geometry, the spins are arranged on a chain, and in the two-dimensional geometry on a square lattice. In both cases, the interaction strengths are parameterized as
\begin{equation}
    J_{pl}=\frac{K_{pl}}{r_{pl}^3},
    \label{eq:J_distance}
\end{equation}
where $r_{pl}$ denotes the distance between spins $p$ and $l$ in lattice units. The spin-spin coupling strengths $K_{pl}$ are sampled uniformly from $[0,20]\,\mathrm{Hz}$, consistent with typical coupling strengths~\cite{Levitt_08, Keeler_10, Clayden_12, Hosur_22}. As a third case, we consider an infinite-dimensional geometry without spatial structure, in which no distance-dependent decay is applied,
\begin{equation}
    J_{pl}=K_{pl}.
    \label{eq:J_infinite}
\end{equation}

For concreteness, we further consider two values of the broadening parameter, $\eta=3\,\mathrm{Hz}$ and $\eta=1\,\mathrm{Hz}$. For each choice of $L$, geometry, and broadening parameter $\eta$, we study a wide range of numbers of shots $R$. Further details about the simulation are given in Appendix~\ref{app:ising_numerics}.

Fig.~\ref{fig:result}(a) and (b) show the infidelity $\mathcal{I}(R,L)$ as a function of the number of shots $R$ for the one-dimensional geometry with broadening $\eta=3\,\mathrm{Hz}$, for $L=10$ and $L=50$ respectively. The scaling $\mathcal{I}(R,L) \sim R^{-1}$ is evident in accordance with Eq.~\eqref{eq:infidelity_scaling}. We find good agreement with this scaling for all considered broadening parameters, interaction geometries and system sizes. Our goal is to determine the scaling function $\mathcal{A}(L)$. Since $\mathcal{I}(R,L)$ is itself a fluctuating quantity, estimated from a finite number of sampled Hamiltonian parameter sets $\theta_\nu$ and shot-noise realizations $\mu$, any estimate of $\mathcal{A}(L)$ extracted from it fluctuates as well. From the large-$R$ scaling $\mathcal{I}(R,L) \approx \mathcal{A}(L)/R$, each fixed value of $R$ provides an
estimate of the scaling function,
\begin{equation}
    A_R(L) \equiv R\,\mathcal{I}(R,L),
    \label{eq:A_R}
\end{equation}
where the dependence on geometry and broadening parameter is kept fixed and left implicit. These estimates are not equally reliable: for smaller $R$, the infidelity $\mathcal{I}(R,L)$ and thus the estimate $A_R(L)$ is more strongly affected by shot noise than for larger $R$. We therefore combine the estimates from different $R$ using an inverse-variance weighted mean, in which an estimate with larger variance receives a smaller weight. Assigning the weight $w_R = 1/\mathrm{Var}[A_R(L)]$ to each value of $R$, the scaling function is
\begin{equation}
    \mathcal{A}(L)
    = \frac{\sum_R w_R\, A_R(L)}{\sum_R w_R}.
    \label{eq:A_weighted}
\end{equation}
The variance $\mathrm{Var}[\mathcal{I}(R,L)]$ is the variance of the sample mean, Eq.~\eqref{eq:infidelity}, obtained from the spread of the single-run infidelities $\mathcal{I}^\mu_\nu(R)$ over the sampled Hamiltonian parameter sets $\theta_\nu$ and shot-noise realizations $\mu$,
\begin{equation}
  \mathrm{Var}[\mathcal{I}(R,L)] = \frac{1}{N_\nu N_\mu}
  \mathrm{Var}_{\nu,\mu}\!\left[\mathcal{I}^\mu_\nu(R)\right].
  \label{eq:var_F}
\end{equation}
The variance of $A_R(L)$ then follows by error propagation,
\begin{equation}
    \mathrm{Var}[A_R(L)] = R^2\,\mathrm{Var}[\mathcal{I}(R,L)].
    \label{eq:var_A_R}
\end{equation}

\begin{figure*}[t!]
 \centering
 \includegraphics[width = 16cm]{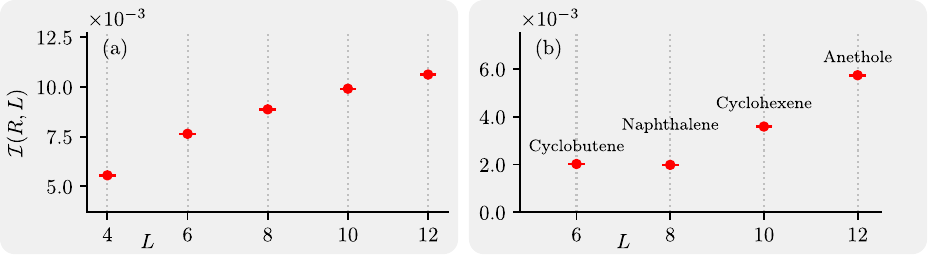}
 \caption{\textbf{Scaling of the infidelity for Heisenberg Hamiltonians.}
    (a) Infidelity $\mathcal{I}(R,L)$ as a function of system size $L$ for the one-dimensional geometry with broadening parameter $\eta=3\,\mathrm{Hz}$ and $R=1024$ shots. The infidelity increases only weakly with system size and does not exhibit a pronounced growth over the considered range of $L$, suggesting that the favorable shot noise behavior observed for the Ising model is not qualitatively changed by the more general Heisenberg interaction, although the data should be understood as an illustrative small-system check rather than a scaling analysis.
    (b) Infidelity $\mathcal{I}(R,L)$ for four example molecules with Heisenberg couplings, evaluated from noisy spectra obtained with $R=1024$ shots and compared to exact reference spectra with a field strength of $20\,\mathrm{MHz}$ and a broadening parameter of $\eta=3\,\mathrm{Hz}$. The infidelity does not show a steeply increasing trend with $L$. Error bars (standard error of the mean) are smaller than the symbols in panels (a) and (b).}
 \label{fig:heisenberg_hamil}
\end{figure*}

The resulting scaling function $\mathcal{A}(L)$ is shown in Fig.~\ref{fig:result}(c)--(h). Across all broadenings and geometries, $\mathcal{A}(L)$ increases with system size, indicating that larger spin systems require more shots to reach the same target infidelity. However, particularly for the one- and two-dimensional geometries this increase becomes consistently weaker with larger $L$, leading to a pronounced flattening of $\mathcal{A}(L)$. Consequently, over the system sizes considered here, the required number of shots grows substantially slower than exponentially.

This behavior may be understood from the intricacy of the NMR spectrum, which affects how $\mathcal{A}(L)$ scales with system size. A spectrum consisting of many peaks that are well separated relative to their linewidth requires each of these peaks to be resolved individually, which demands high accuracy in the time signal and therefore many shots; the corresponding $\mathcal{A}(L)$ is large. Frequencies that lie close together relative to their linewidth, by contrast, contribute little to our resource criterion [Eq.~\eqref{eq:cosine_sim}], since it measures the overlap of the full noisy spectrum with the reference spectrum rather than the resolution of each individual frequency. The intricacy of the spectrum in this sense is controlled by the system size $L$, the interaction geometry, and the broadening parameter $\eta$. 

In the following we discuss how the intricacy increases or decreases as the geometry and $\eta$ are varied. Increasing $L$ raises the number of possible transitions between states of the many-body spin system, and thus the number of peaks in the NMR spectrum. How strongly each added spin affects the spectrum depends on the geometry. In the infinite-dimensional geometry, every spin couples to all others, so the number of relevant couplings per spin keeps growing with $L$. Each added spin thus contributes further resolvable frequencies to the spectrum, and reproducing them to a fixed accuracy requires more shots. This is consistent with the strong growth of $\mathcal{A}(L)$ observed for this geometry. In the one- and two-dimensional geometries, by contrast, the $1/r_{pl}^3$ decay suppresses the influence of distant spins, so the additional splittings are small compared to the spacing of the resonances set by the local Zeeman coefficients $h_l$. The corresponding peaks therefore lie close to the resonance positions already present in the spectrum, and reproducing them adds little to the accuracy, and hence to the number of shots, required to reach a given infidelity. This offers a possible explanation for the flattening of $\mathcal{A}(L)$ observed in these geometries. 

The broadening $\eta$ sets the linewidth of the individual peaks. We do not observe a qualitative difference in the scaling function between broadening parameter $\eta = 1\,\mathrm{Hz}$ and $\eta = 3\,\mathrm{Hz}$.

\subsection{Heisenberg interactions}
\label{sec:heisenberg_interaction}
We now repeat the same analysis for NMR molecules with Heisenberg interactions, Eq.~\eqref{eq:heisenberg_hamiltonian}. In this case the exact dynamics can no longer be simulated efficiently, therefore we restrict our analysis to one-dimensional systems with up to $L=12$ spins.

We use the same one-dimensional parameter ensemble as in the Ising case, i.e.\ sampling the Hamiltonian parameters $\theta = \{h_l, J_{pl}\}$ from the previously defined intervals [cf. Sec.~\ref{sec:ising_interaction}]. For each sampled Hamiltonian, we generate the noisy spectra by simulating the repeated-measurement protocol of Sec.~\ref{sec:shot_noise} explicitly: we sample initial product eigenstates $\ket{x_j}$ of $\hat{M}^x$, evolve them using exact diagonalization of the Heisenberg Hamiltonian, and perform projective measurements of $\hat{M}^x$ and $\hat{M}^y$. Averaging over the shots gives the estimators $\overline{M}^x_{R,\theta}(t_n)$ and $\overline{M}^y_{R,\theta}(t_n)$ [cf. Eq.~\eqref{eq:finite_shot_estimator}]. The noisy spectra are then obtained as before and compared to exact reference spectra [see Appendix~\ref{app:heisenberg_numerics} for the numerical details].

The resulting infidelity is shown in Fig.~\ref{fig:heisenberg_hamil}(a). Since we use only a single, fixed number of shots, we do not extract $\mathcal{A}(L)$ here; instead we use that $\mathcal{I}(R,L) \approx \mathcal{A}(L)/R$, so at fixed $R$ the $L$-dependence of $\mathcal{I}(R,L)$ reflects that of $\mathcal{A}(L)$. The infidelity increases only weakly with $L$, indicating that the favorable scaling found for the Ising model is not qualitatively altered by the Heisenberg interaction.
Since the Heisenberg interaction, Eq.~\eqref{eq:heisenberg_hamiltonian}, describes the couplings in real NMR molecules, we test our findings on actual molecules. In Fig.~\ref{fig:heisenberg_hamil}(b), we consider four examples, cyclobutene, naphthalene, cyclohexene, and anethole. Also here, the infidelity does not increase steeply with $L$, suggesting that our findings carry over to real molecules.


\section{Conclusion and Outlook}
In this work we have investigated the scaling of the number of shots required to obtain NMR spectra with a desired accuracy in a low field setting. We focused on shot noise, a fundamental issue even for error corrected quantum computing, and neglected other possible error sources such as gate noise, decoherence, and time-evolution approximation errors. We first employed a simplified model of the NMR Hamiltonian, in which we replaced the Heisenberg interaction term by Ising interactions. Across different interaction geometries and broadening parameters, we found a flattening of the required number of shots with the system size. The flattening is more pronounced for interaction geometries in which interactions are suppressed with the distance between individual spins. These results indicate that shot noise is not a prohibitive obstacle for NMR spectrum simulation with digital quantum computers, even for large molecules with more than 25 NMR-active spins. This supports the possible usefulness of digital quantum computers for larger NMR systems. An important next step was to investigate whether the same favorable behavior persists for more realistic Heisenberg NMR Hamiltonians. The small-system checks indicate that the qualitative shot noise behavior is not drastically changed, but they do not yet establish the scaling with system size. At least for the system sizes and interaction geometries considered here, our results suggest that the number of shots required for NMR spectrum simulation remains far below the exponential growth of the underlying Hilbert space. These findings help support the notion that the simulation of NMR spectra, as well as the dynamics of quantum spin systems more generally, is an encouraging use case for quantum computing.

\acknowledgments
We acknowledge funding from the Baden-Württemberg Ministry of Science, Research and Arts through the IQST project "Defining, quantifying and assessing a practical quantum advantage in the computation of NMR spectra", as well as the German Federal Ministry of Education and Research, through project QSolid (13N16155). Support was also received through the ERC grant OPEN-2QS (Grant No. 101164443).

\noindent\textit{Data availability}---The data and code that support the findings of this article are openly available~\cite{zenodo}.

\noindent\textit{AI usage}---The authors acknowledge using Anthropic's Claude Opus 5.5 for refactoring the authors' code for publication. All suggestions were reviewed and approved by the authors.

\bibliography{sample.bib}

\clearpage
\appendix
\onecolumngrid

\section{Mean and variance of the single-shot estimator}
\label{app:estimator}

Here we compute the mean and variance of the single-shot estimator. For one shot $r$, the protocol of Sec.~\ref{sec:shot_noise} yields the product $\tilde{M}^{\alpha}_{r,\theta}(t) = x_j \alpha_i$ of the initial and final magnetization eigenvalues. Each shot yields one realization of this random variable. Its mean equals the exact magnetization expectation value $\langle \hat{M}^{\alpha}_{\theta}(t)\rangle$ of Eq.~\eqref{eq:correlation_function_gen}, and its variance, which we denote by
\begin{equation}
    v^{\alpha}(t) \equiv \mathrm{Var}\!\left[\tilde{M}^{\alpha}(t)\right]
\end{equation}
and refer to as the \emph{single-shot variance}, quantifies the statistical fluctuations of a single measurement outcome around this mean. By the central limit theorem, the $R$-shot average $\overline{M}^{\alpha}_{R,\theta}(t)$ of Eq.~\eqref{eq:finite_shot_estimator} is, for large $R$, normally distributed around $\langle \hat{M}^{\alpha}_{\theta}(t) \rangle$ with variance $v^{\alpha}(t)/R$.

In the remainder of this Appendix, we derive explicit expressions for $\langle \hat{M}^{\alpha}_{\theta}(t) \rangle$ and $v^{\alpha}(t)$ for the Ising Hamiltonian, Eq.~\eqref{eq:ising_hamiltonian}. Throughout, we fix one Hamiltonian realization and suppress the parameter label $\theta$ for notational simplicity.
\subsection*{Magnetization expectation values for Ising interactions}
The expectation values, Eq.~\eqref{eq:correlation_function_gen}, are
\begin{equation}
    \langle \hat{M}^\alpha(t)\rangle
    =\frac{1}{2^L}
    \mathrm{Tr}\!\left[
        \hat{M}^\alpha(t)\hat{M}^x
    \right],
    \qquad
    \hat{M}^\alpha(t)
    =
    e^{+2\pi i\hat{H} t}
    \hat{M}^\alpha
    e^{-2\pi i\hat{H} t},
\end{equation}
with $\alpha\in\{x,y\}$ and $\hat{M}^\alpha=\sum_i\hat{I}_i^\alpha$. Since the Ising Hamiltonian is diagonal in the $z$ basis, the Heisenberg evolution of a transverse operator $\hat{I}_i^\alpha$ is governed only by the terms containing spin $i$; all other terms commute with $\hat{I}_i^\alpha$ and drop out. These relevant terms are
\begin{equation}
    \hat{H}_{i}
    =
    \hat{I}_i^z \hat{\Omega}_i,
    \qquad
    \hat{\Omega}_i
    =
    h_i+\sum_{j\neq i}J_{ij}\hat{I}_j^z ,
\end{equation}
where $\hat{\Omega}_i$ acts only on the spectator spins $j\neq i$ and has eigenvalues $\Omega_i(\{s_j\}) = h_i+\sum_{j\neq i}J_{ij}s_j$ on the spectator configurations $\ket{\{s_j\}_{j\neq i}}$, with $s_j=\pm\tfrac12$. The transverse operators then evolve as 
\begin{align}
    \hat{I}_i^x(t)
    &=
    \hat{I}_i^x\cos(2\pi\hat{\Omega}_i t)
    -
    \hat{I}_i^y\sin(2\pi\hat{\Omega}_i t),
    \\
    \hat{I}_i^y(t)
    &=
    \hat{I}_i^y\cos(2\pi\hat{\Omega}_i t)
    +
    \hat{I}_i^x\sin(2\pi\hat{\Omega}_i t).
\end{align}

Inserting $\hat{M}^\alpha(t)=\sum_i\hat{I}_i^\alpha(t)$ and $\hat{M}^x=\sum_k\hat{I}_k^x$ into the trace, only the equal-site terms $i=k$ survive, since a single transverse operator on any site traces to zero. For $\alpha=x$, using $(\hat{I}_i^x)^2=\tfrac14\mathbb{I}_i$, $\hat{I}_i^y\hat{I}_i^x=-\tfrac{i}{2}\hat{I}_i^z$, and $\mathrm{Tr}_i[\hat{I}_i^z]=0$, the sine term drops out and
\begin{equation}
    \mathrm{Tr}\!\left[\hat{M}^x(t)\hat{M}^x\right]
    =
    \frac{1}{2}
    \sum_i
    \mathrm{Tr}_{j\neq i}\!\left[
        \cos(2\pi\hat{\Omega}_i t)
    \right].
\end{equation}
The spectator trace is a sum over all independent configurations $s_j=\pm\tfrac12$. Writing the cosine as the real part of an exponential, this sum factorizes over the spectator spins,
\begin{equation}
    \mathrm{Tr}_{j\neq i}\!\left[
        \cos(2\pi\hat{\Omega}_i t)
    \right]
    =
    \mathrm{Re}\left[
        e^{2\pi i h_i t}
        \prod_{j\neq i}
        \sum_{s_j=\pm 1/2}
        e^{2\pi i J_{ij}s_j t}
    \right]
    =
    2^{L-1}
    \cos(2\pi h_i t)
    \prod_{j\neq i}
    \cos(\pi J_{ij}t).
\end{equation}
The corresponding expectation value is therefore
\begin{equation}
     \langle \hat{M}^x(t)\rangle
     =
    \frac{1}{4}
    \sum_{i=1}^{L}
    \cos(2\pi h_i t)
    \prod_{j\neq i}
    \cos(\pi J_{ij}t).
    \label{eq:expectation_x}
\end{equation}
For $\alpha=y$, the same steps retain the sine term instead, giving
\begin{equation}
    \langle \hat{M}^y(t)\rangle
    =
    \frac{1}{4}\sum_{i=1}^{L}
    \sin(2\pi h_i t)
    \prod_{j\neq i}
    \cos(\pi J_{ij}t).
\end{equation}

\subsection*{Single-shot variance for Ising interactions}
We now compute the single-shot variance $v^{x}(t)$. In the following, $a$ and $b$ label many-body basis states of the estimator $\tilde{M}^{x}(t) = x_a x_b$, while $i,j,k$ are reserved for spin sites. With the mean given by Eq.~\eqref{eq:expectation_x}, the variance
\begin{equation}
    v^{x}(t)
    = \mathbb{E}\!\left[(\tilde{M}^{x})^{2}\right]\!(t)
    - \mathbb{E}\!\left[\tilde{M}^{x}\right]\!(t)^{2}
\end{equation}
requires the second moment
\begin{equation}
    \mathbb{E}\!\left[(\tilde{M}^{x})^{2}\right]\!(t)
    = \frac{1}{2^{L}}
      \operatorname{Tr}\!\left[(\hat{M}^{x})^{2}(\hat{M}^{x}(t))^{2}\right].
\end{equation}
Expanding both factors with $(\hat{I}_i^x)^2=\tfrac14\mathbb{I}_i$,
\begin{equation}
    (\hat M^x)^2
    =
    \frac{L}{4}\mathbb I
    +
    2\sum_{i<j}\hat I_i^x\hat I_j^x ,
    \qquad
    (\hat M^x(t))^2
    =
    \frac{L}{4}\mathbb I
    +
    2\sum_{p<q}\hat I_p^x(t)\hat I_q^x(t),
\end{equation}
and taking the normalized trace, the constant term gives $L^2/16$, while the terms linear in a single pair vanish because $\mathrm{Tr}[\hat I_i^x\hat I_j^x]=0$ (and likewise for the evolved pair, using invariance of the trace under time evolution). Only the product of one unevolved and one evolved pair remains,
\begin{equation}
    \mathbb{E}\!\left[(\tilde M^x)^2\right](t)
    =
    \frac{L^2}{16}
    +
    4
    \sum_{i<j}\sum_{p<q}
    \frac{1}{2^L}
    \mathrm{Tr}\!\left[
        \hat I_i^x\hat I_j^x
        \hat I_p^x(t)\hat I_q^x(t)
    \right].
\end{equation}
The full trace factorizes into local traces, each of which vanishes unless the operators on that site multiply to the identity. The unevolved pair carries transverse operators on sites $i,j$, and the evolved pair on sites $p,q$; a nonzero trace therefore requires the transverse supports to coincide, i.e. $\{i,j\}=\{p,q\}$. The double sum collapses to
\begin{equation}
    \mathbb{E}\!\left[(\tilde M^x)^2\right](t)
    =
    \frac{L^2}{16}
    +
    4
    \sum_{i<j}
    T_{ij}(t),
    \qquad
    T_{ij}(t)
    =
    \frac{1}{2^L}
    \mathrm{Tr}\!\left[
        \hat I_i^x\hat I_j^x
        \hat I_i^x(t)\hat I_j^x(t)
    \right].
\end{equation}

To evaluate $T_{ij}(t)$, we fix the $z$-eigenvalues $s_k=\pm\tfrac12$ of all spectator spins $k\neq i,j$ and sum over them at the end. For a fixed spectator configuration, the Hamiltonian on the two active spins is, up to spectator-only constants,
\begin{equation}
    \hat H_{ij}
    =
    a_i\hat I_i^z
    +
    a_j\hat I_j^z
    +
    J_{ij}\hat I_i^z\hat I_j^z,
    \qquad
    a_{i}
    =
    h_{i}+\sum_{k\neq i,j}J_{ik}s_k ,
\end{equation}
and analogously for $a_j$. In the two-spin $z$-basis $\ket{z_i,z_j}$ with energies $E(z_i,z_j)=a_i z_i+a_j z_j+J_{ij}z_i z_j$, the operator $\hat I_i^x\hat I_j^x$ flips both spins, $\hat I_i^x\hat I_j^x\ket{z_i,z_j}=\tfrac14\ket{-z_i,-z_j}$, so the diagonal matrix element entering the trace is $\tfrac{1}{16}e^{2\pi i[E(-z_i,-z_j)-E(z_i,z_j)]t}$. The interaction $J_{ij}$ cancels in the energy difference,
\begin{equation}
    E(-z_i,-z_j)-E(z_i,z_j)
    =
    -2a_i z_i-2a_j z_j ,
\end{equation}
because flipping both spins leaves $z_i z_j$ unchanged. Summing over the four states with $\sum_{z=\pm1/2}e^{-4\pi i a z t}=2\cos(2\pi a t)$ gives $\tfrac14\cos(2\pi a_i t)\cos(2\pi a_j t)$. Including the normalized sum over the $2^{L-2}$ spectator configurations,
\begin{equation}
    T_{ij}(t)
    =
    \frac{1}{16}
    \frac{1}{2^{L-2}}
    \sum_{\{s_k=\pm1/2\}}
    \cos(2\pi a_i t)\cos(2\pi a_j t).
\end{equation}
Applying $\cos A\cos B=\tfrac12[\cos(A+B)+\cos(A-B)]$ with $a_i\pm a_j = h_i\pm h_j+\sum_{k\neq i,j}(J_{ik}\pm J_{jk})s_k$, the spectator sums factorize as before, yielding $T_{ij}(t)=\tfrac{1}{16}B_{ij}(t)$ with
\begin{equation}
    \begin{split}
    B_{ij}(t)
    =
    \frac12\Bigg[
    &
    \cos(2\pi(h_i+h_j)t)
    \prod_{k\neq i,j}
    \cos\!\left(\pi(J_{ik}+J_{jk})t\right)
    \\
    &+
    \cos(2\pi(h_i-h_j)t)
    \prod_{k\neq i,j}
    \cos\!\left(\pi(J_{ik}-J_{jk})t\right)
    \Bigg].
    \end{split}
\end{equation}
The single-shot variance is therefore
\begin{equation}
    v^{x}(t)
    =
    \frac{L^2}{16}
    +
    \frac14
    \sum_{i<j}
    B_{ij}(t)
    -
    \left[
        \frac14
        \sum_{i=1}^{L}
        \cos(2\pi h_i t)
        \prod_{j\neq i}
        \cos(\pi J_{ij}t)
    \right]^2 .
\end{equation}
The calculation for the $y$ component is analogous, with the initial state sampled from the local $x$-basis and the final measurement in the local $y$-basis. With $\tilde M^y(t)=x_a y_b$, one obtains
\begin{equation}
     v^{y}(t)
    =
    \frac{L^2}{16}
    +
    \frac14
    \sum_{i<j}
    B^{(y)}_{ij}(t)
    -
    \left[
        \frac14
        \sum_{i=1}^{L}
        \sin(2\pi h_i t)
        \prod_{j\neq i}
        \cos(\pi J_{ij}t)
    \right]^2 ,
\end{equation}
where
\begin{equation}
    \begin{split}
    B^{(y)}_{ij}(t)
    =
    \frac12\Bigg[
    &
    \cos(2\pi(h_i-h_j)t)
    \prod_{k\neq i,j}
    \cos\!\left(\pi(J_{ik}-J_{jk})t\right)
    \\
    &-
    \cos(2\pi(h_i+h_j)t)
    \prod_{k\neq i,j}
    \cos\!\left(\pi(J_{ik}+J_{jk})t\right)
    \Bigg].
    \end{split}
\end{equation}

\section{Large-$R$ scaling of the infidelity}
\label{app:scaling}

We derive the large-$R$ scaling of the infidelity used in Eq.~\eqref{eq:infidelity_scaling}. Throughout, we fix one Hamiltonian realization and suppress the parameter label $\theta$. As established in Appendix~\ref{app:estimator}, the estimate $\overline{M}^{\alpha}_{R}(t_n)$ at each time point $t_n$ and direction $\alpha = x,y$ fluctuates around the exact expectation value $\langle \hat{M}^{\alpha}(t_n)\rangle$ with variance $v^{\alpha}(t_n)/R$. We therefore write
\begin{equation}
    \overline{M}^{\alpha}_{R}(t_n)
    = \langle \hat{M}^{\alpha}(t_n)\rangle + \xi^{\alpha}_{R}(t_n),
\end{equation}
where the finite-shot fluctuation satisfies
\begin{equation}
    \mathbb{E}[\xi^{\alpha}_{R}(t_n)] = 0,
    \qquad
    \mathrm{Var}[\xi^{\alpha}_{R}(t_n)] = \frac{v^{\alpha}(t_n)}{R}.
\end{equation}
We now collect the finite-shot estimates at all time points and both directions into vectors. The exact magnetization values form
\begin{equation}
    \mathbf{m}
    =
    \left(
    \langle \hat{M}^x(t_0)\rangle,\ldots,\langle \hat{M}^x(t_{N-1})\rangle,
    \langle \hat{M}^y(t_0)\rangle,\ldots,\langle \hat{M}^y(t_{N-1})\rangle
    \right)^T ,
\end{equation}
and the finite-shot fluctuations form
\begin{equation}
    \boldsymbol{\xi}_R
    =
    \left(
    \xi^x_R(t_0),\ldots,\xi^x_R(t_{N-1}),
    \xi^y_R(t_0),\ldots,\xi^y_R(t_{N-1})
    \right)^T .
\end{equation}
The vector of finite-shot magnetization estimates is then their sum,
\begin{equation}
    \overline{\mathbf{m}}_R = \mathbf{m} + \boldsymbol{\xi}_R .
\end{equation}

The finite-shot estimates of the signal are obtained independently for each time point $t_n$ and for each measurement direction $\alpha=x,y$. Therefore, the fluctuations $\xi^\alpha_R(t_n)$ are uncorrelated between different times and directions, and the covariance matrix of $\boldsymbol{\xi}_R$ is diagonal,
\begin{equation}
\mathrm{Cov}\!\left[\boldsymbol{\xi}_{R}\right]
=
\frac{1}{R}\Sigma_{t},
\end{equation}
with
\begin{equation}
\Sigma_{t}
=
\mathrm{diag}\!\left(
v^x(t_0),\ldots,v^x(t_{N-1}),
v^y(t_0),\ldots,v^y(t_{N-1})
\right).
\end{equation}
For large $R$, the central limit theorem implies that $\boldsymbol{\xi}_{R}$ is approximately Gaussian. The important point for the following scaling argument is that its covariance is proportional to $1/R$.

The spectrum is obtained from the magnetization time signal by a linear transformation: the $x$ and $y$ components are combined into the complex FID, multiplied by the exponential decay factor, and Fourier transformed, keeping the real part [cf.\ Eqs.~\eqref{eq:FID}, \eqref{eq:spectrum}, and \eqref{eq:shot_noise_spectra}]. Since the real part is taken, this defines a real linear map $B \in \mathbb{R}^{N \times 2N}$ acting on the stacked vector $\overline{\mathbf{m}}_R$, whose elements combine the $x$ and $y$ components into the real spectrum. Applying this map to the finite-shot magnetization vector gives the finite-shot spectral vector
\begin{equation}
    \mathbf{s}_R = B\,\overline{\mathbf{m}}_R
    = B\,\mathbf{m} + B\,\boldsymbol{\xi}_R .
\end{equation}

We identify
\begin{equation}
\mathbf s
=
B\mathbf m,
\qquad
\boldsymbol{\epsilon}_{R}
=
B\boldsymbol{\xi}_{R},
\end{equation}
where $\mathbf s$ is the exact spectral vector and $\boldsymbol{\epsilon}_{R}$ is the spectral noise induced by the finite-shot fluctuations.

Because the transformation is linear, the covariance of the spectral noise is
\begin{equation}
\mathrm{Cov}\!\left[\boldsymbol{\epsilon}_{R}\right]
=
B\,\mathrm{Cov}\!\left[\boldsymbol{\xi}_{R}\right]B^T
=
\frac{1}{R}B\Sigma_t B^T
\equiv
\frac{1}{R}\Sigma_f .
\end{equation}
Thus, although the spectral noise is generally correlated between different frequencies, its covariance still scales as $1/R$.

We now expand the cosine similarity between the noisy spectrum $\overline{\mathbf s}_{R}=\mathbf s+\boldsymbol{\epsilon}_{R}$ and the exact spectrum $\mathbf s$ in powers of the finite-shot noise $\boldsymbol{\epsilon}_{R}$. Starting from
\begin{equation}
C(\overline{\mathbf s}_{R},\mathbf s)
=
\frac{
(\mathbf s+\boldsymbol{\epsilon}_{R})^T\mathbf s
}{
\|\mathbf s+\boldsymbol{\epsilon}_{R}\|\,\|\mathbf s\|
},
\end{equation}
we define
\begin{equation}
a=\mathbf s^T\mathbf s,
\qquad
b=\mathbf s^T\boldsymbol{\epsilon}_{R},
\qquad
c=\boldsymbol{\epsilon}_{R}^T\boldsymbol{\epsilon}_{R}.
\end{equation}
Here, $b$ is first order in the noise, while $c$ is second order. The cosine similarity can then be written as
\begin{equation}
C(\overline{\mathbf s}_{R},\mathbf s)
=
\frac{a+b}{\sqrt{a}\sqrt{a+2b+c}}
=
\frac{1+b/a}{\sqrt{1+2b/a+c/a}}.
\end{equation}
Using
\begin{equation}
(1+x)^{-1/2}
=
1-\frac{x}{2}+\frac{3x^2}{8}
+O(x^3),
\end{equation}
with $x=2b/a+c/a$, and keeping terms up to second order in $\boldsymbol{\epsilon}_{R}$, we obtain
\begin{align}
C(\overline{\mathbf s}_{R},\mathbf s)
&=
\left(1+\frac{b}{a}\right)
\left[
1-\frac{1}{2}\left(\frac{2b}{a}+\frac{c}{a}\right)
+\frac{3}{8}\left(\frac{2b}{a}\right)^2
\right]
+O(\boldsymbol{\epsilon}_{R}^3)
\\
&=
1
-\frac{1}{2}\frac{c}{a}
+\frac{1}{2}\frac{b^2}{a^2}
+O(\boldsymbol{\epsilon}_{R}^3).
\end{align}
Consequently,
\begin{equation}
1-C(\overline{\mathbf s}_{R},\mathbf s)
\approx
\frac{1}{2}
\frac{\boldsymbol{\epsilon}_{R}^T\boldsymbol{\epsilon}_{R}}
{\|\mathbf s\|^2}
-
\frac{1}{2}
\frac{
(\mathbf s^T\boldsymbol{\epsilon}_{R})^2
}
{\|\mathbf s\|^4}.
\end{equation}

The leading contribution to $1-C(\overline{\mathbf s}_{R},\mathbf s)$ is quadratic in the finite-shot spectral noise. To see its $R$-dependence, we take the expectation value over the random measurement outcomes for the fixed Hamiltonian. Since the finite-shot fluctuations have zero mean,
\begin{equation}
\mathbb E[\boldsymbol{\xi}_R]=0,
\end{equation}
and since $\boldsymbol{\epsilon}_R=B\boldsymbol{\xi}_R$, the spectral noise also has zero mean,
\begin{equation}
\mathbb E[\boldsymbol{\epsilon}_R]=0.
\end{equation}
Therefore, its covariance matrix is
\begin{equation}
\mathrm{Cov}\!\left[\boldsymbol{\epsilon}_R\right]
=
\mathbb E\!\left[
\boldsymbol{\epsilon}_R\boldsymbol{\epsilon}_R^T
\right]
=
\frac{1}{R}\Sigma_f .
\end{equation}
Using this, we obtain
\begin{equation}
\mathbb E\!\left[
\boldsymbol{\epsilon}_R^T\boldsymbol{\epsilon}_R
\right]
=
\mathrm{tr}\!\left(
\mathrm{Cov}\!\left[\boldsymbol{\epsilon}_R\right]
\right)
=
\frac{1}{R}\mathrm{tr}(\Sigma_f)
\end{equation}
and
\begin{equation}
\mathbb E\!\left[
(\mathbf s^T\boldsymbol{\epsilon}_R)^2
\right]
=
\mathbf s^T
\mathrm{Cov}\!\left[\boldsymbol{\epsilon}_R\right]
\mathbf s
=
\frac{1}{R}\mathbf s^T\Sigma_f\mathbf s .
\end{equation}
Both expectation values entering the leading expression for $\mathbb E[1-C(\overline{\mathbf s}_{R},\mathbf s)]$ therefore scale as $1/R$. Reintroducing the Hamiltonian label $\theta$, this gives, for each fixed Hamiltonian realization,
\begin{equation}
\mathbb E\!\left[
1-C(S_{R,\theta},S_\theta)
\right]
\sim
\frac{1}{R}.
\end{equation}
The infidelity $\mathcal I(R,L)$ used in the main text is obtained by averaging single-run infidelities over the sampled Hamiltonian realizations. Since this averaging changes only the prefactor, but not the $R$-dependence, one obtains
\begin{equation}
\mathcal I(R,L)
\approx
\frac{\mathcal A(L)}{R}.
\end{equation}

\section{Numerical details for the analysis of the Ising Hamiltonian}
\label{app:ising_numerics}
Here we provide the numerical parameters used to generate the Ising-model data of Fig.~\ref{fig:result}, complementing the description in Sec.~\ref{sec:ising_interaction}. For each sampled Hamiltonian, the noisy spectra are obtained by computing the exact expectation value, Eq.~\eqref{eq:correlation_function_gen}, and the single-shot variances $v^\alpha(t_n)$ [cf. Appendix~\ref{app:estimator}] in direction $\alpha =x $ and $\alpha = y$ for the given Hamiltonian parameters. We obtain the noisy estimates of the time signal (Eq.~\eqref{eq:finite_shot_estimator}) by drawing from the corresponding normal distribution, as described in Sec.~\ref{sec:ising_interaction}.

For each system size $L$ and geometry, we draw $N_\nu = 1000$ Hamiltonians from the ensemble defined in the main text and, for each, generate $N_\mu = 1000$ independent noisy spectra. We do this for the system sizes $L = 10, 15, 20, \dots, 50$, and for each system size we use $14$ shot numbers between $R = 2.5 \times 10^4$ and $R = 10^6$.

The time signal is sampled on an equidistant grid of $N = 16384$ points over an evolution time $T = 10\,\mathrm{s}$, shared by both broadenings so that the two datasets are directly comparable. The evolution time is long enough that the FID has decayed to a negligible level by $t = T$, and the resulting Nyquist frequency of $819\,\mathrm{Hz}$ lies well above the spectral support for all considered parameters, so that neither truncation nor aliasing artefacts affect the spectra.

The exact reference spectra are likewise obtained from the exact expectation value evaluated on the same discrete time grid, rather than from the closed-form continuous-time NMR spectrum available for the Ising interaction \cite{Levitt_08, Keeler_10, Kuprov_11}. This ensures that reference and noisy spectra share an identical discretization, so that any deviation between them is attributable solely to shot noise.

\section{Numerical details for the analysis of the Heisenberg Hamiltonian}
\label{app:heisenberg_numerics}

Here we provide the numerical parameters used to generate the Heisenberg-model data of Fig.~\ref{fig:heisenberg_hamil}, complementing the description in Sec.~\ref{sec:heisenberg_interaction}. As before, the exact reference spectra are obtained from the exact expectation value, Eq.~\eqref{eq:correlation_function_gen}, evaluated on the same discrete time grid as the noisy spectra.

The time signal is sampled on an equidistant grid of $N = 8192$ points over an evolution time $T = 3.0\,\mathrm{s}$.

For the randomly sampled Hamiltonians, we draw $400$ independent Hamiltonians for each system size $L = 4,6,8,10,12$ and generate $1000$ noisy spectra per Hamiltonian, with $R = 1024$ shots and broadening parameter $\eta = 3\,\mathrm{Hz}$.

The Hamiltonian parameters for Naphthalene are obtained from Hans Reich's Collection - NMR Spectroscopy~\cite{Hans_Reich}.
The Hamiltonian parameters of the other three molecules (Cyclobutene, Cyclohexene and Anethole) are obtained from the HQS Spectrum Tools NMR parameter database~\cite{hqspectrum_25}. The $J$-couplings are taken directly in Hz, while the local Zeeman coefficients are obtained from the tabulated chemical shifts $\delta_l$ (in ppm) via $h_l = \delta_l \nu_0$, at a spectrometer frequency of $\nu_0 = 20\,\mathrm{MHz}$. We generate $1000$ noisy spectra per molecule, with $R = 1024$ shots and broadening $\eta = 3\,\mathrm{Hz}$.

\clearpage
\newpage

\twocolumngrid

\setcounter{equation}{0}
\setcounter{figure}{0}
\renewcommand{\theequation}{A\arabic{equation}}
\renewcommand{\thefigure}{A\arabic{figure}}

\onecolumngrid
\newpage

\setcounter{equation}{0}
\setcounter{page}{1}

\setcounter{figure}{0}
\setcounter{table}{0}
\makeatletter
\renewcommand{\theequation}{S\arabic{equation}}
\renewcommand{\thefigure}{S\arabic{figure}}
\renewcommand{\thetable}{S\arabic{table}}
\setcounter{secnumdepth}{1}

{
	\thispagestyle{plain}
	\clearpage
}

\end{document}